\documentclass[prl,twocolumn,showpacs]{revtex4}

\usepackage[applemac]{inputenc}
\usepackage[T1]{fontenc}
\usepackage{ae}
\usepackage{graphicx} 
\usepackage{amsfonts} 
\usepackage{amssymb} 
\usepackage{amsmath} 
\usepackage{latexsym} 
\usepackage{amsthm} 
\usepackage{booktabs} 
\usepackage{psfrag} 
\usepackage{float} 
\usepackage{color} 
\usepackage{paralist} 
\usepackage{caption} 
\usepackage{subfigure} 
\usepackage{hyperref} 

\newcommand{\beq}{\begin{equation}}
\newcommand{\eeq}{\end{equation}}

\def\ket#1{\mathinner{|{#1}\rangle}}

\providecommand{\abs}[1]{\lvert#1\rvert}

\begin{document}

\title{Universal Energy Distribution of Quasiparticles \\ Emitted in a Local Time-Dependent Quench.}

\author{Pietro Smacchia${}^\dag$ and Alessandro Silva${}^\ddag$}

\address{$^\dag$ SISSA, International School for Advanced Studies, via Bonomea 265, 34136 Trieste, Italy}
\address{${}^\ddag$ICTP , International Centre for Theoretical Physics, P.O. Box 586, 34014 Trieste, Italy}

\begin{abstract}
We study the emission of quasi-particles in the scaling limit of the $1$D Quantum Ising chain at the critical point perturbed by a time dependent local transverse field. We  compute \it exactly \rm and for a \it generic \rm time dependence the average value of the transverse magnetization, its correlation functions, as well as the statistic of both the inclusive and exclusive work. We show that, except for a cyclic perturbation, the probability distribution of the work at low energies  is a power law whose exponent is universal, i.e. does not depend on the specific time dependent protocol, but only on the final value attained by the perturbation.
\end{abstract}

\pacs{05.70.Ln, 05.30.-d, 02.30.Ik}

\maketitle

The study of out-of-equilibrium dynamics of many-body quantum systems has gained  renewed interest over the past decade. This is mainly a result of impressive experimental advances in the field of cold atoms \cite{cold_atoms}, exemplified by the observation of the collapse and revival of a system driven across the Mott-superfluid transition \cite{collapse_revival}. In many cases a common way of taking a system out of equilibrium is the so called \textit{quantum quench}, i.e. a variation in time of a parameter $g$ of the Hamiltonian describing the system. The resulting dynamical response can be probed in  different ways, i.e. by looking at the evolution of correlation functions \cite{Calabrese_Cardy} or entanglement entropies \cite{entropy_cft,entropy2}, or  adopting a  thermodynamical point of view and studying the statistic of the work done on the system \cite{work}, the change in entropy \cite{Polkovnikov} or the energy distribution after the quench \cite{Polkovnikov2}.

Quantum quenches can be either \it global \rm or \it local \rm in space. In both cases a change of the system parameters causes the emission of quasiparticles carrying correlations that travel across the system with a certain velocity $v$, whose maximum value for a quantum many-body system is given by the Lieb-Robinson bound \cite{Lieb_Robinson}. The difference between the two cases lies in the fact that while in the global scenario the emission happens everywhere,  in the local case it is restricted to the point where the quench is performed, behaving as a sort of ``quantum antenna''\cite{antenna}. Qualitatively, the result of this mechanism is the appearance of the so-called ``light-cone'' effect \cite{Calabrese_Cardy}, which has been very recently experimentally observed  in bosonic ultracold gases\cite{Bloch}. In the case of a local quench, on which we will focus in the following, the ligh-cone effect implies that the time lapse to see the effect of a quench at a distance $x$ from the antenna is $t \sim x/v$, the time it takes for excitations to reach that point \cite{Everts}. 
The effects of a quench are particularly strong at a critical point where the excitation spectrum is gapless. In this case, a conformal field theory (CFT) describes well a large class of $1$d quantum systems for sufficiently large distances and time scales (see for example \cite{CFT}).

Independently on the protocol considered, the semiclassical picture above suggests that a local quench will in general produce a correlation front propagating at the velocity $v$ of the quasiparticles emitted, without reference to either the energy distribution of the excitations created or to the profile of the average values of physical observables. These quantities are  however important to
characterize how much energy is transferred to the system by the local perturbation and what is the form of the signal that propagates through the sample. These issues, which are in turn relevant for applications of time dependent quantum protocols in quantum information \cite{antenna, Cardy_local} and quantum optimization problems \cite{optimization}, were hardly addressed in the literature where most of the local quenches considered are ``abrupt''; i.e., the parameter is suddenly changed from an initial value $g_0$ to a final one $g_f$ \cite{Gallavotti}. 

In this Letter we consider a \textit{generic time dependent} local quench of the transverse field in a quantum Ising chain at the critical point. Describing the system in its scaling limit,
we solve \it exactly \rm the dynamical problem, computing the average transverse magnetization, its correlations and the statistics of the work done for a {\it generic protocol}. In doing so, we show
that the behavior of the probability distribution function of the work at low energies is a power law whose exponent does not depend on the detailed time dependence of the protocol considered, but just on the final value of the local transverse field (with the exception of the case in which this value is zero). We finally comment on the extension of these results to more general systems.

Before entering the technical details of our analysis we will summarize and discuss the main results obtained. 
We will consider a quantum Ising chain, $H=-1/2 \sum_i\sigma^x_i\sigma^x_{i+1}+g\sigma_i^z$, where $\sigma_i^{x,z}$ are the longitudinal and 
transverse spin operators at site $i$ and $g$ is the strength of the transverse field. Setting the global transverse field $g=1$, the system is at its quantum critical point:
here the mass $m= 1-g$ of quasiparticle excitations vanishes and quasiparticles propagate freely through the system at the "speed of light" $v=1$. 
In the following, we will be interested in studying the physics of a local change  of the transverse field away from $g=1$. In order to analyze this problem, we
describe the system in the scaling limit by its corresponding CFT~\cite{Mussardo},  locally perturbed by a mass term
\beq
H_t= -\frac{i}{2} \int dx \left[ \varphi \partial_x \varphi- \bar{\varphi} \partial_x \bar{\varphi} \right]+ i m(t) \bar{\varphi} \varphi_{\vert_{x=0}},
\label{eq:ham}
\eeq
where $\varphi$ and $\bar{\varphi}$ are two Majorana fermionic operators, so that $\left\{\varphi(x),\varphi(x')\right\}=\left\{\bar{\varphi}(x), \bar{\varphi}(x')\right\}=\delta(x-x')$, and $m(t)=0$ for all $t\leq t_0$, with $t_0$ arbitrary initial time, in such a way as to have the system  in its ground state until $t=t_0$. 
Let us start by characterizing, for a generic time-dependent protocol, the energy distribution of quasiparticles emitted. In order to do so, imagine performing two measurements of energy, one before and one after the quench. The resulting energy difference is the so-called inclusive work $w$ done on the system~\cite{Talkner_rev}, which, for a system out of equilibrium,  is a stochastic quantity characterized by a  probability distribution $P_i(w)$ \cite{Jarzynski, Kurchan, Talkner_rev}.  We have computed its characteristic function  $\mathcal{G}_i(u)=\int dw \, e^{i u w} P_i (w)$ for a generic quench protocol, obtaining 
\beq
\begin{split}
\mathcal{G}_i(u)=\exp\bigg[&\frac{1}{4 \pi^2} \int_{-\infty}^\tau \!\!\!\!dt  \int_{-\infty}^\tau dt'  \partial_t m(t) \partial_{t'} m(t')\\ &\log\frac{\alpha-i (t-t')}{\alpha-i (t-t'+u)}\bigg],
\end{split}
\label{eq:inc_g}
\eeq
where $\tau$ is the time at which the final energy measurement is performed and $\alpha$ is the ultraviolet cut-off of the theory. The work done on the system depends on the derivative with respect to time of the protocol $m(t)$ chosen, matching with the expectation that this quantity should be related to the speed at which the quench is performed.

Let us now show that the form of $P_i(w)$ for small $w$, giving us information of the energy transmitted to the system, is \it independent \rm on the specifics of the protocol employed, i.e. universal. For this sake, we have to analyze the asymptotics of $\mathcal{G}_i$ for large $u$ (much larger than the time scale of the protocol) . When $m(\tau) \neq 0$ we get $\mathcal{G}_i (u) \sim (-i u )^{-m(\tau)^2/4 \pi^2}$, corresponding to
\beq
P_i(w) \overset{w \rightarrow 0}{\sim} w^{\frac{m(\tau)^2}{4 \pi^2}-1}.
\eeq
As anticipated, the distribution $P_i(w)$ displays an edge singularity at small $w$ with an exponent \it independent \rm on the details of the protocol $m(t)$ chosen but just on its amplitude. In particular for small quenches ($m(\tau)< 2 \pi$) there is a power law divergence, while for large quenches $P_i(w)$ vanishes with a cusp. This  matches with the natural physical expectation that the more the parameter in the Hamiltonian is changed, the smaller the probability of doing very small work on the system is. 

\begin{figure}
\begin{center}
\subfigure[\label{fig:ex_w1}]{\includegraphics[width=4.2 cm, height=3.3 cm]{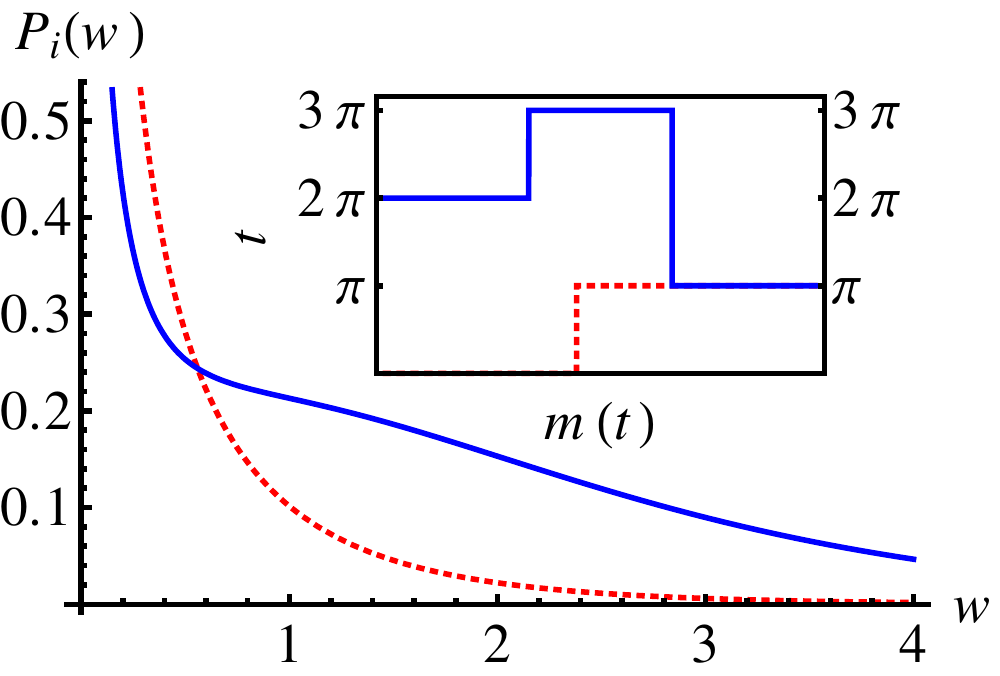}}
\subfigure[\label{fig:ex_w2}]{\includegraphics[width=4.3 cm, height=3.3 cm]{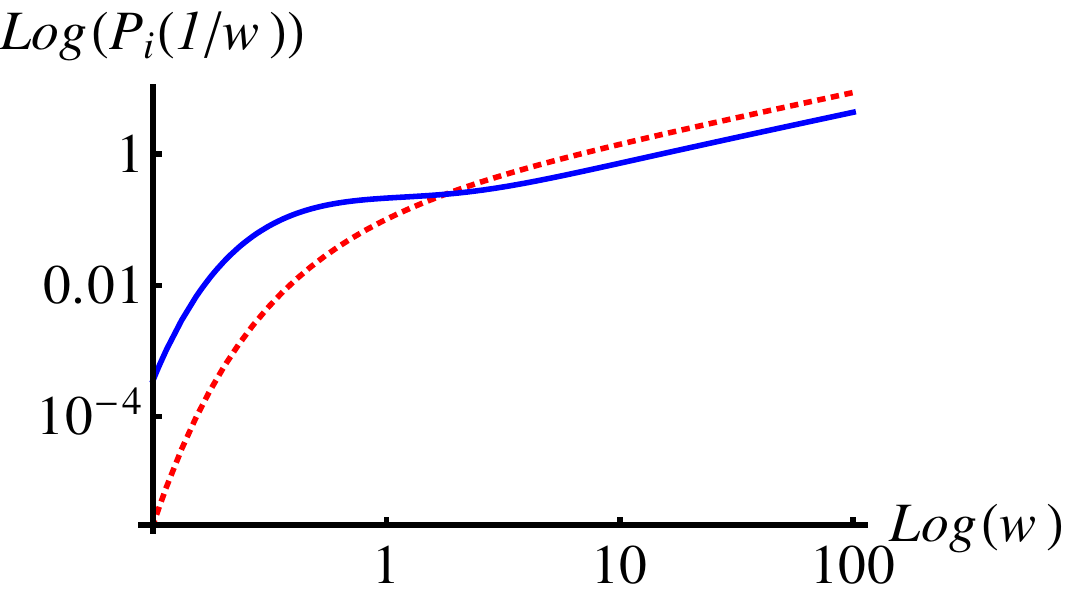}}
\caption{\footnotesize (Color online) (a) Probability distributions $P_i(w)$ for a nonmonotonic protocol (Blue,Full), i.e., a series of sudden quenches and a sudden quench (Red, Dashed) ending at the same value of m and shown in the inset. (b) Logarithmic plot of $P_i(1/w)$ for the same protocols as before. We take $\alpha=1$.}
\label{fig:example_work}
\end{center}
\end{figure}

On one hand the independence of the low energy behavior of the distribution of the work from the details of the protocol can be a natural expectation in the case of monotonic protocols, since when $u$ is large they all look like sudden quenches, making the details of how the final value of $m$ is reached irrelevant. The result we obtained is, however, more general: it holds independently on the shape of the protocol, no matter what happens before the end, and therefore even in cases of nonmonotonic protocols, where the former similarity is not true any more. We also note that, in contrast with the case of global quenches, where in the thermodynamic limit the spectral weight of the distribution is concentrated in a peak at high energies, the low energy part of the distribution of the work can retain a considerable spectral weigth. This means that the power law behavior can be observable. The example of Fig. \ref{fig:example_work} clarifies both the issue of nonmonotonicity and observability. In Fig. \ref{fig:ex_w1} $P_i(w)$ is shown  for a non monotonic protocol and a sudden quench to the same final value of $m$ (see the inset). One can see that in both cases the low energy part has a considerable spectral weight. From Fig. \ref{fig:ex_w2} instead one can see that the two protocols at low energy indeed behave as a power law with the same exponent.

Cyclic protocols with $m(\tau)=0$ are an exception to the scenario above, since in this case the asymptotic behavior becomes $\mathcal{G}_i(u)\sim\exp\left[k-i\frac{k'}{u^2}\right]$, where $k$ and $k'$ are two constants depending on the specific form of $m(t)$. In this case $P_i(w)$ will  have a delta-function peak, $\delta(w)$ with a nonuniversal amplitude, plus a regular part vanishing linearly.  This means that for each cyclic protocol there is a nonzero probability of not doing work on the system.  The absence of the delta function for finite $m(\tau)$ is in turn a consequence of the Anderson orthogonality catastrophe \cite{Anderson}.

\begin{figure}
\begin{center}
\subfigure[\label{fig:secondexamplea}]{\includegraphics[width=4.2 cm, height=3 cm]{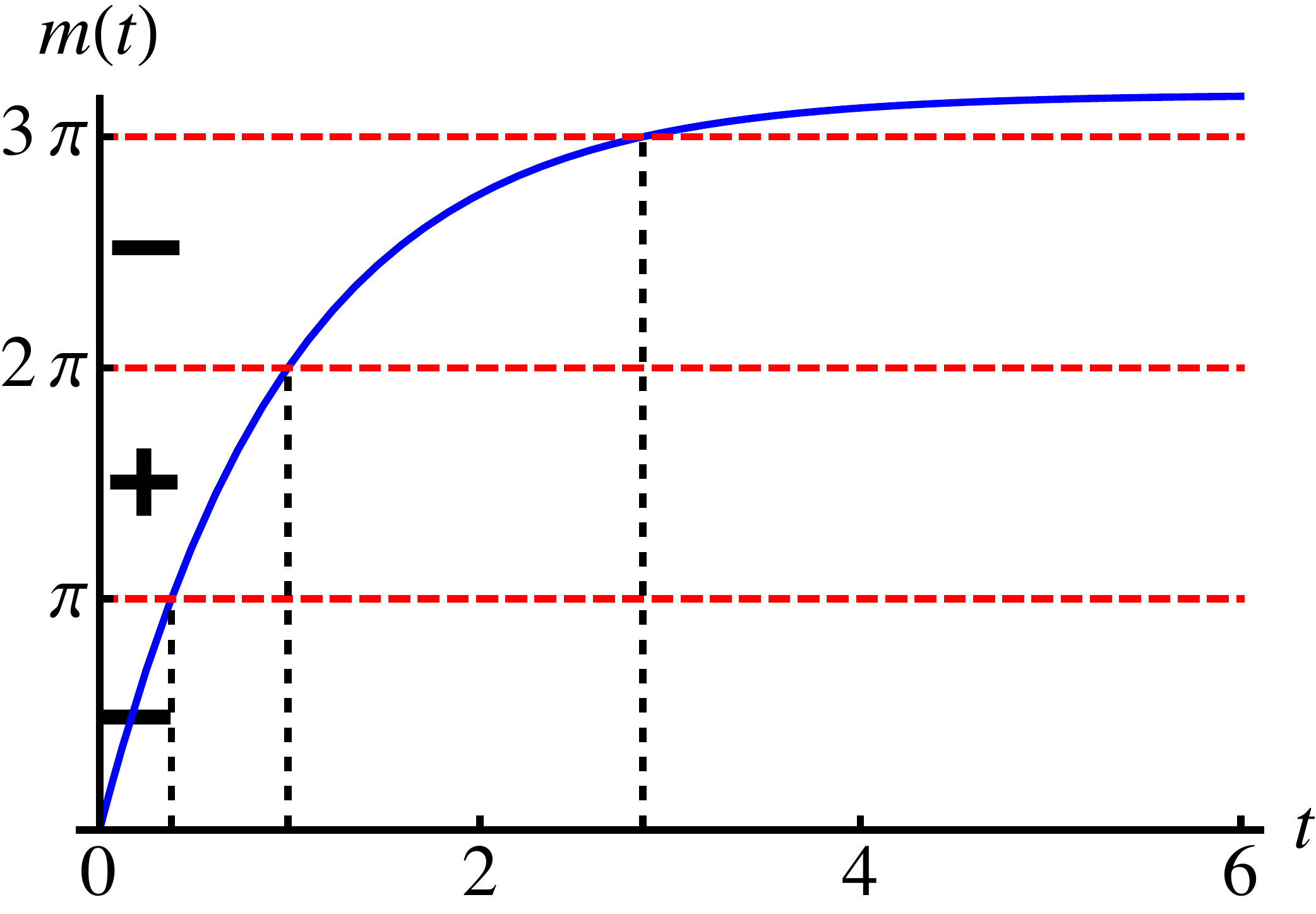}}
\subfigure[\label{fig:secondexampleb}]{\includegraphics[width=4.2 cm, height=3 cm]{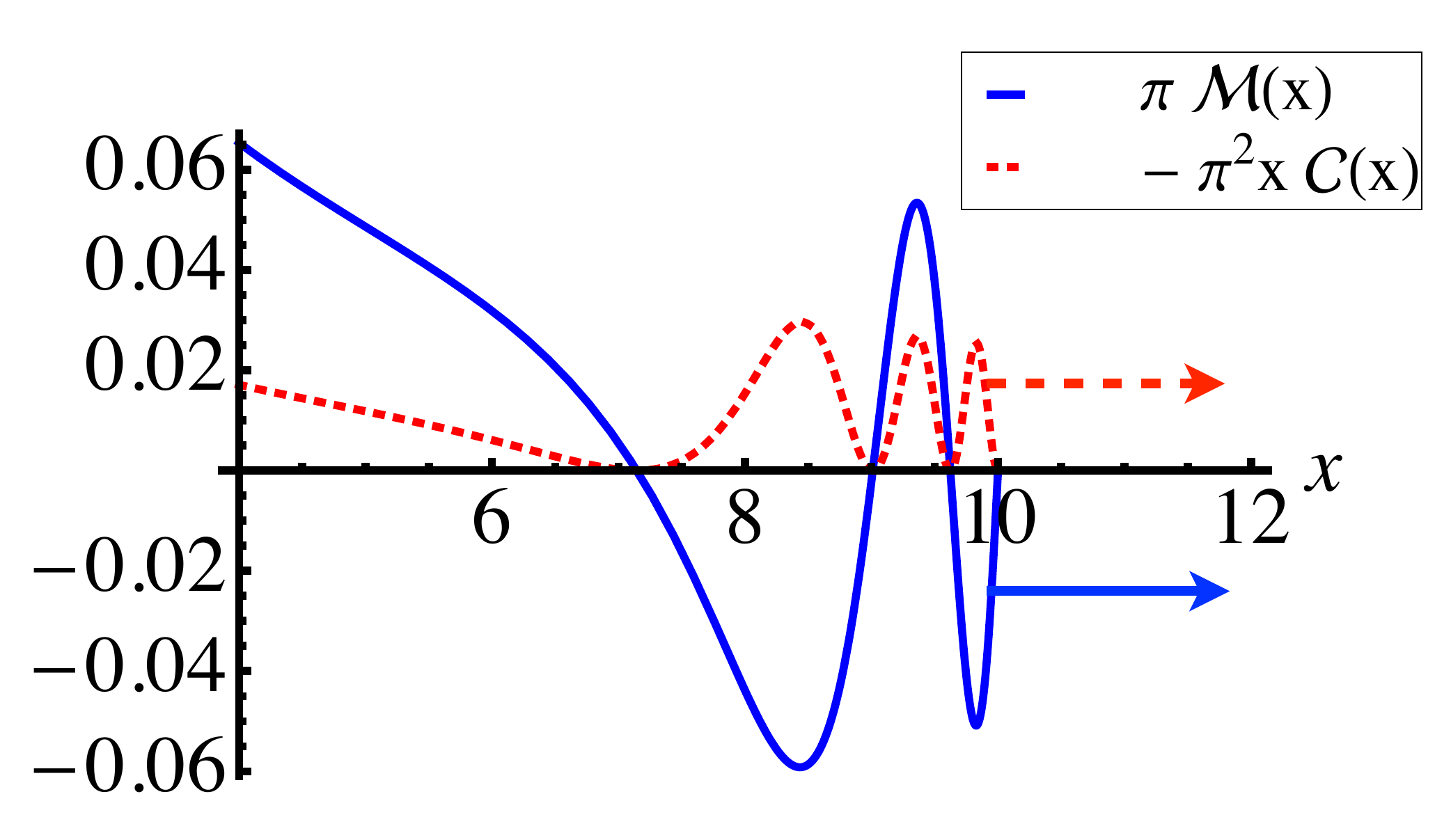}}
\subfigure[\label{fig:thirdexamplea}]{\includegraphics[width=4.2 cm, height=3 cm]{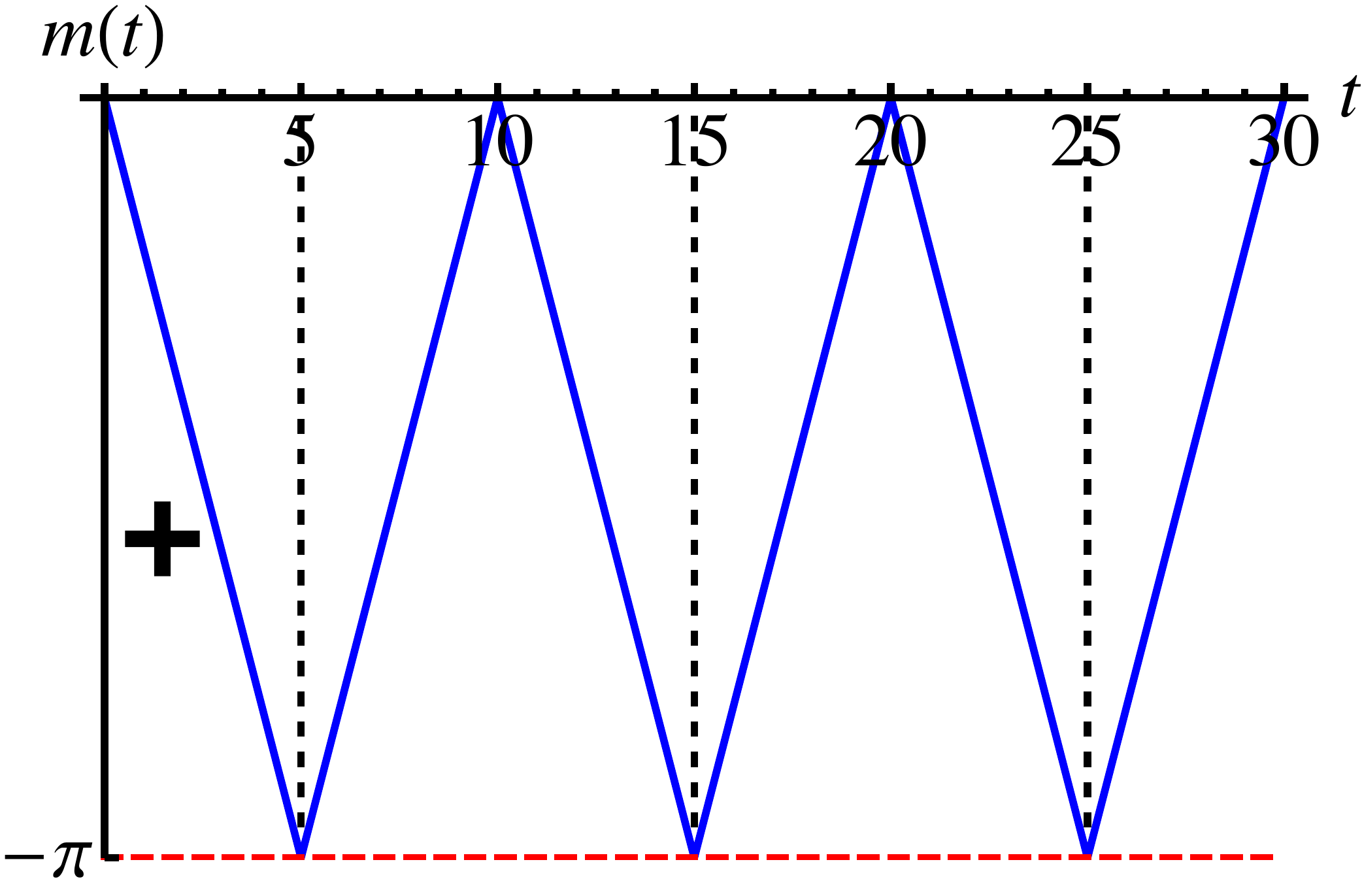}}
\subfigure[\label{fig:thirdexampleb}]{\includegraphics[width=4.2 cm, height=3 cm]{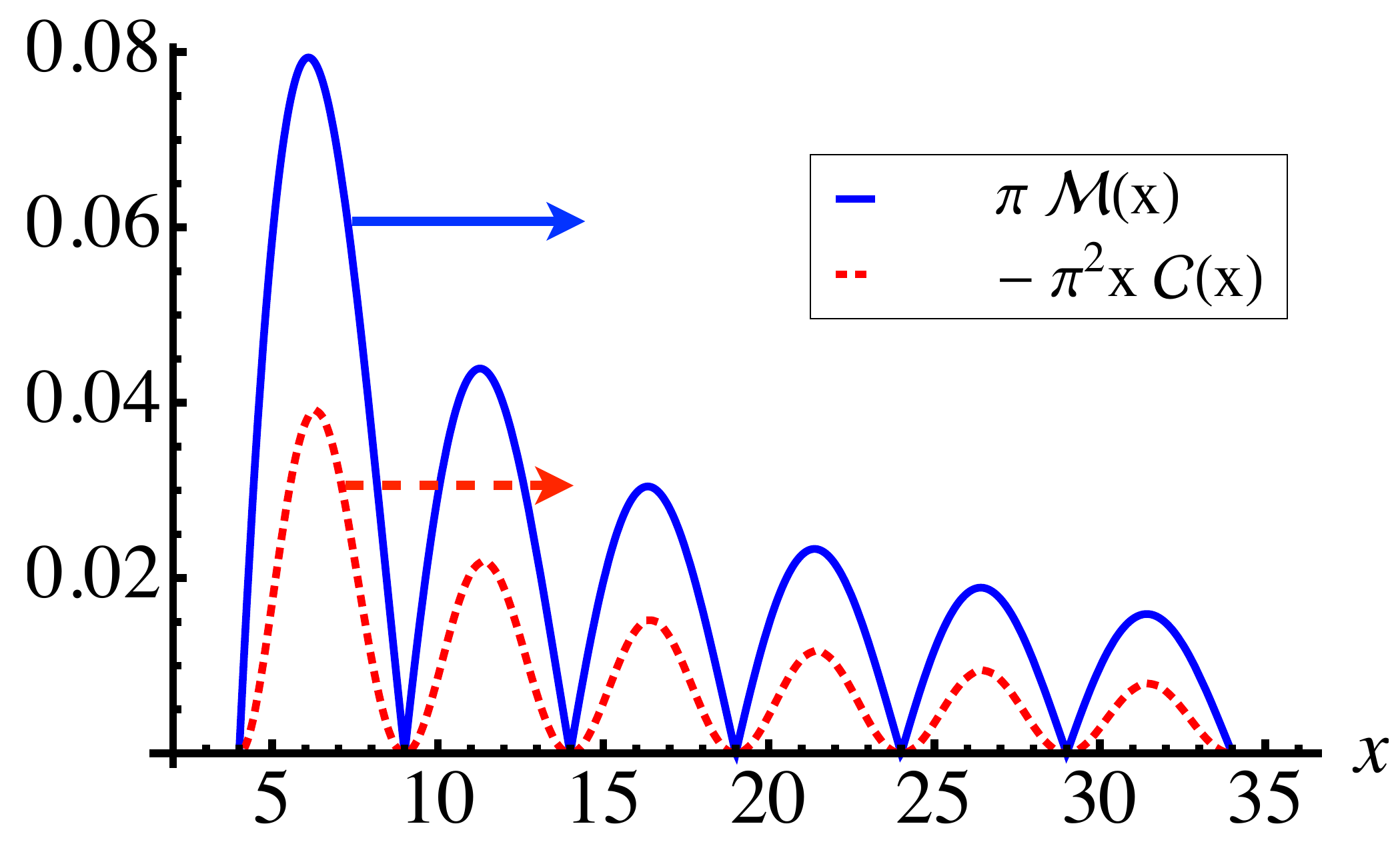}}
\caption{\footnotesize (Color online) Examples of magnetization and correlations profiles (right) for some specific protocol $m(t)$ (left). $t=10$ in (b) and $t=30$ in (d). $\alpha=0$ in all the cases.}
\end{center}
\end{figure}

The possibility to infer the behavior of physical quantities from some gross features of the protocol $m(t)$ is also observed for the transverse magnetization and its correlation function. The result for the average value of the  transverse magnetization $\mathcal{M} (x,t)$, represented in the scaling limit by the operator $2 i\, \bar{\varphi}(x,t) \varphi(x,t)$, is
\beq
\langle \mathcal{M} (x,t) \rangle = - \frac{ 2 \abs{x}}{\pi(4 x^2+\alpha^2)} \sin\left(m(t-\abs{x})\right).
\label{eq:transverse_magn}
\eeq 
We clearly observe that the protocol $m(t)$ leads to the propagation at the velocity of light (which has been taken equal to $1$) of two identical magnetization signals, one to the left and one to the right of the origin. The strength of the signals ,in turn, decreases  with the distance as $1/x$.  The qualitative features of the traveling profile can be easily  extracted. 
For example in Fig. \ref{fig:secondexamplea} we have taken $m(t)= 10 (1-e^{-t}) \Theta(t)$. The traveling profile will have three zeros, since $m(t)$ crosses the values $\pi$, $2 \pi$ and $3 \pi$, and will have a positive tail, since it asymptotically ends at a value between $3 \pi$ and $4 \pi$. One can check from Fig. \ref{fig:secondexampleb} that these are indeed the features of the generated profile. This simple understanding can be used to design protocols $m(t)$ producing a profile with the desired features. For instance we show in Fig. \ref{fig:thirdexamplea} a protocol that produces six positive wave-packets with the same width.

Finally, let us show how to extend this analysis to more complex physical quantities, e.g. the connected correlation function of the transverse magnetization. At equal times this is given by
\begin{widetext}
\beq
\begin{split}
& \langle \mathcal{M}(x,t) \mathcal{M}(x',t) \rangle_C= \cos \left( m(t-\abs{x})\right) \cos\left( m(t-\abs{x'})\right) \left[\frac{1} {2 \pi^2 \left[(x-x')^2+\alpha^2\right]} + \frac{\alpha^2}{2 \pi^2 (4 x^2+\alpha^2) (4x'^2+\alpha^2)} \right]\\
& -\sin \left( m(t-\abs{x})\right) \sin\left( m(t-\abs{x'})\right) \left[\frac{ 4 \abs{x} \abs{x'}}{2 \pi^2 (4 x^2+\alpha^2) (4 x'^2+\alpha^2)} -\frac{ 2 \Theta(x x') x x'}{\pi^2 \left[ (x+x')^2+\alpha^2\right]\left[(x-x')^2+\alpha^2\right]} \right].
\end{split}
\eeq
\end{widetext}
Since we expect quasiparticle to be emitted symmetrically, the correlation between two opposite points $x$ and $-x$ is of particular interest. In particular, the \it excess \rm correlation $\mathcal{C}(x,t)=\langle \mathcal{M}(x,t) \mathcal{M}(-x,t) \rangle_C-\langle \mathcal{M}(x,t_0) \mathcal{M}(-x,t_0) \rangle_C$, is given by
\beq
\mathcal{C}(x,t)=\frac{1}{2 \pi^2 (4 x^2+\alpha^2)} \left( \cos \left(2 m(t-\abs{x}) \right)-1 \right).
\eeq
As before, one may easily design the protocol $m(t)$ to give rise to a certain correlation profile. In particular, the zeroes of $\mathcal{C}(x,t)$  are the same as those of the magnetization, as one can see from Figs. \ref{fig:secondexampleb} and \ref{fig:thirdexampleb} for specific protocols.
More specifically, for every protocol $m(t)$ the excess correlations are always negative and travel through the systems at the same speed as the magnetization, decreasing with the distance from the origin as $1/x^2$.  

Let us now discuss the \it exact \rm solution of this dynamical problem  as well as the method used to obtain the results given above. In doing so we will also discuss in more detail the work done on the system, extending the results  of Eq. (\ref{eq:inc_g}) to the exclusive work. Let us first sketch the main steps of the analysis. For every physical quantity at hand, it will be consistently useful to compute it by duplicating our theory~\cite{Itzykson}, i.e., introducing an additional couple of Majorana fermions $\chi$, $\bar{\chi}$ also described by the Hamiltonian of Eq. (\ref{eq:ham}). Two pairs of Majorana fermions can be indeed used to form two Dirac fermions $\psi_R= e^{-i \pi/4} (\varphi+i \chi)/\sqrt{2}$ and $\psi_L=e^{i \pi/4} (\bar{\varphi}+i \bar{\chi}) \sqrt{2}$, which can be combined trough a nonlocal transformation \cite{Kane}, defining $\psi_+(x)=(\psi_R(x)+\psi_L(-x))/\sqrt{2}$ and $\psi_-(x)=(\psi_R(x)-\psi_L(-x))/\sqrt{2}i$, which are described by the Hamiltonian
\beq
H_t=i\!\!\int \!\!\!dx \left[ \psi^\dagger_- \partial_x\psi_-  - \psi^\dagger_+ \partial_x \psi_+ \right]+m(t) \left[ \psi^\dagger_+ \psi_+ - \psi^\dagger_- \psi_- \right]_{\vert_{x=0}}.
\eeq 

This Hamiltonian describes two decoupled chiral modes  completely characterized by the single particle Hamiltonians $H_{+,-}= \mp i \partial_x \pm \delta(x) m(t)$. The latter gives for the fields the equations of motion
$\left[i \partial_t\pm i \partial_x \right]\psi_{+,-}(x,t)= \pm \delta(x) m(t) \psi_{+,-}(x,t)$,
with the initial condition that $\psi_{+,-}(x,t_0)$ are free fermionic operators.
These equations describe the scattering of chiral modes on a time-dependent delta potential and their solutions can be written as 
\beq
\psi_{+,-}(x,t)= e^{\mp i m(t-\abs{x}) \theta(\pm x)} \psi_{+,-}(x\mp t,t_0).
\label{eq:pm_evolution}
\eeq

The computation of  the average value of the magnetization and of its correlation functions proceeds now by expressing the operator $2i\, \bar{\varphi} \varphi= i \bar{\varphi} \varphi+i \bar{\chi} \chi$ in terms of the $\psi_+$ and $\psi_-$ operators, by applying the transformations given above, and then by using the mode expansion of the fermionic operators, taking into account that the average is taken on the Dirac sea of both fermions $+$ and $-$ (all modes with negative momentum occupied, all states with positive momentum free). The same procedure can be applied to compute the correlations.

The probability distribution of the work can in turn be computed by considering that for each realization of an out-of-equilibrium protocol the work $w$ is given as a difference of the outcomes of two measures of the energy, at the initial time $t=t_0$ and at the final time $t= \tau$ \cite{Talkner_rev}. The final energy is measured with respect to the full Hamiltonian $H_\tau$, or to the initial Hamiltonian $H_{t_0}$, excluding the forcing term, depending on whether one is interested in the \it inclusive \rm or \it exclusive \rm work done. For example, in the inclusive case we have $P_i(w)= \sum_{n,m} \delta \left(w-\left(E_n(\tau)-E_m(t_0)\right) \right) p(n|m,\tau) p_m $,
with $p(n|m,\tau)= \lvert \langle \psi_n(\tau)|U(\tau,t_0)|\psi_m(t_0)\rangle \rvert^2$, and $p_m=\lvert \langle \psi_m(t_0)|\Phi(t_0)\rangle\rvert$, where $\ket{\Phi(t_0)}$ is the initial state of the system, $U(t,t_0)$ is the evolution operator from $t_0$ to $t$, and the equation $H_t |\psi_i(t)\rangle= E_i(t) |\psi_i(t)\rangle$ is valid. In the exclusive case the definition of $P_e(w)$ is the same if $E_n(\tau)$ and $\psi_n(\tau)$ are replaced by $E_n(t_0)$ and $\psi_n(t_0)$. 

As shown in Ref.~\onlinecite{Talkner_rev}, the characteristic functions $\mathcal{G}_{i,e}(u)=\int dw \, e^{i u w} P_{i,e} (w)$ contain full information about the statistics of $w$ and can be written as a two-time correlations function,
\beq
\mathcal{G}_{i,e}(u)=\langle e^{i u H^{H}_{\tau,t_0}} e^{-i u H_{t_0}}\rangle,
\label{eq:char_func}
\eeq
where $H^H_{\tau,t_0}= U^\dagger(\tau,t_0) H_{\tau,t_0} U(\tau,t_0)$ is the Hamiltonian used in the second measurement in the Heisenberg representation at time $\tau$.

To compute such quantities we use again the trick due to Itzykson and Zuber \cite{Itzykson} which consists in considering $\mathcal{G}_{i,e}(u)^2$. In order to do so we bosonize the Heisenberg representation at time $\tau$ of the Hamiltonian (\ref{eq:pm_evolution}) with the usual formula $\psi_{\pm}(x,t)= 1/ \sqrt{2 \pi \alpha} \, e^{\pm i \sqrt{4 \pi} \phi_{\pm}(x,t)}$, getting
\beq
\begin{split}
H^H_\tau=&\int dx \, \left(\partial_x \phi_+(x,\tau) \right)^2+\left( \partial_x \phi_-(x,\tau) \right)^2+\\
& \frac{m(\tau)}{\sqrt{\pi}} \left(\partial_x \phi_+-\partial_x \phi_{-} \right)_{\vert_{x=0,\tau}}.
\end{split}
\eeq
Then we observe that, apart from an irrelevant constant, we can write $H^H_\tau=\mathcal{U}^{\dagger}_iH_{t_0} \mathcal{U}_i$, or for the sake of computing the exclusive work, $H^H_0=\mathcal{U}^{\dagger}_eH_{t_0} \mathcal{U}_e$, with $\mathcal{U}_i=e^{i/\sqrt{\pi} \hat{A}}$, $ \mathcal{U}_e= e^{i/\sqrt{\pi} \hat{B}}$, where

\beq
\hat{A}= \int_0^\infty dy \left( \phi_+(y-\tau,0)+\phi_-(\tau-y,0) \right) \partial_y m(\tau-y),
\eeq
and
\beq
\hat{B}= \hat{A}+m(\tau) \left(\phi_+(-\tau,0)+\phi_-(\tau,0) \right).
\eeq
Then from Eq. (\ref{eq:char_func}) we get 
\beq
\mathcal{G}_{i,e}^2=\langle \mathcal{U}_{i,e}^\dagger e^{i u H_{t_0}}\mathcal{U}_{i,e}e^{-i u H_{t_0}}\rangle= \langle \mathcal{U}_{i,e}^\dagger \mathcal{U}_{i,e}(u) \rangle,  
\label{eq:formula_g}
\eeq
where the average is taken on the bosonic vacuum and $\mathcal{U}_{i,e}(u)$ means that the bosonic operators are evolved at time $u$ with the free bosonic Hamiltonian $H_{t_0}$. Eq. (\ref{eq:formula_g}) can then be computed with standard methods.

While the result for the statistics of the \it inclusive \rm work has been previously anticipated, for the \it exclusive \rm work we obtain
\beq
\begin{split}
&\mathcal{G}_e(u)=\mathcal{G}_i(u)\times \exp\bigg[\frac{1}{4 \pi^2} \left(m(\tau)^2 \log \frac{\alpha}{\alpha-i u} \right.\\
& \left. -m(\tau) \int_{-\infty}^\tau \partial_t m(t) \log \frac{\alpha^2+(\tau-t)^2}{(\alpha-i u)^2+(\tau-t)^2} \right)\bigg].
\end{split}
\label{eq:exc_g}
\eeq

When the final value of the mass $m(\tau)$ is zero the inclusive (Eq. \ref{eq:inc_g}) and exclusive (Eq. \ref{eq:exc_g}) characteristic functions coincide.
Let us now briefly analyze the asymptotic behavior of $\mathcal{G}_e (u)$ when $m(\tau)\neq 0$. One obtains in general $\mathcal{G}_e \sim\exp\left[k_1-i\frac{k_1'}{u^2}\right] $, with $k_1$ and $k'_1$ depending on $m(t)$, implying that $P_e(w)$  consists of a delta function at zero as well as a regular part linearly vanishing as $w \rightarrow 0$.

Finally we briefly discuss the possibility of extending the result we found for $P_i(w)$ at low energy. Given that a power law should be expected from the orthogonality catastrophe, if we make an analytic continuation $u \rightarrow i R$ in Eq. (\ref{eq:char_func}), we can interpret the characteristic function as a partition function of the classical correspondent model on a strip of thickness $R$, with a line defect stretching between the two  boundary states $U(\tau,t_0) \ket{\Phi(t_0)}$. The behavior at large $R$, that will determine the behavior at small $w$ of $P_i(w)$, is expected to depend on the renormalization group flow of the boundary states as well as  of the final Hamiltonian. For a local quench, we expect the boundary state to flow back to the critical one, since a non-extensive number of defects has been generated by the protocol. On the other hand,  the defect could be relevant, marginal or irrelevant. In the case of a marginal defect (as the one explicitly considered above) the exponent should depend on the final strength of the defect, while for a relevant perturbation we do expect the exponent to be completely independent on the quench performed and equal to $c/8-1$, where $c$ is the central charge, coming from the effect of a line of defect in a generic CFT \cite{Cardy_local} . A confirmation of this last statement can be observed in the case of sudden quenches (i.e., the Fermi edge problem) in Luttinger liquids \cite{Luttinger}.

In conclusion, we characterized the signal propagating through the system and the energy transmitted in a  generic time-dependent local quench of the transverse field in a quantum critical system, the quantum Ising chain at the critical point.  By solving exactly the problem in the scaling limit, we have shown that the work done on the system at low energies does not depend on the details of the protocol, but just on its amplitude. The independence of the physics on the fine details of the protocol $m(t)$ is observed also in the traveling signal of the magnetization profile and on its correlations.

\it Acknowledgements \rm - We would like to thank G. Santoro, R. Fazio and J. Cardy for useful discussions and the Galileo Galilei Institute for Theoretical Physics for the hospitality during the completion of this work.

\end{document}